\documentclass[aps,prb,twocolumn,groupedaddress,amsmath,amssymb]{revtex4}

\usepackage{amssymb}
\usepackage{amsmath}
\usepackage{graphicx}
\usepackage{subfigure}
\usepackage{textcomp}
\usepackage{color}
\usepackage{amsfonts}
\usepackage{bbold}
\usepackage{dsfont}
\usepackage{epsfig}
\usepackage{hyperref}

\newcommand{\arctanh}[1]{\operatorname{arctan}}

\begin{document}

\title{Simulating STM transport in alkanes from first principles} 
\date{\today} 
\author{C. Toher\footnote{Current address: Dresden University of Technology, Dresden, Germany} and S. Sanvito}
\email{sanvitos@tcd.ie}
\affiliation{School of Physics and CRANN, Trinity College, Dublin 2, Ireland}

\begin{abstract}
Simulations of scanning tunneling microscopy measurements for molecules on surfaces are traditionally 
based on a perturbative approach, most typically employing the Tersoff-Hamann method. This assumes that the 
STM tip is far from the sample so that the two do not interact with each other. However, when the tip gets close 
to the molecule to perform measurements, the electrostatic interplay between the tip and substrate may generate non-trivial potential 
distribution, charge transfer and forces, all of which may alter the electronic and physical structure of the molecule. 
These effects are investigated with the {\it ab initio} quantum transport code {\sc smeagol}, 
combining non-equilibrium Green's functions formalism with density functional theory. In particular, we investigate 
alkanethiol molecules terminated with either CH$_3$ or CF$_3$ end-groups on gold surfaces, for which recent 
experimental data are available. We discuss the effects connected to the interaction between the STM tip and the 
molecule, as well as the asymmetric charge transfer between the molecule and the electrodes.
\end{abstract}


\maketitle

\section{Introduction}

Scanning tunnelling microscopy (STM) \cite{STM} is an invaluable surface characterization tool with multiple 
applications in molecular electronics. Typically, an atomically sharp probing tip constructed from heavy metals 
such as platinum, tungsten or iridium scans a substrate by means of a tiny tunneling current. The method can be 
used as a topographic tool to map the positions of atoms, defects and molecules on surfaces, as a spectroscopic tool 
to probe their local density of states, and more recently in its spin-polarized version as an ultra sensitive magnetometer 
\cite{Wisen}. Furthermore STM can also be used as an atomic fabrication tool for depositing atoms and molecules 
on a surface so as to form nanoscale devices \cite{coral}. Additionally, it can be used to manufacture breaking junctions, 
where a molecule is pulled out of a surface, so that electronic transport measurements can be performed \cite{Tao1, Tao2}. Indeed, the
level of geometric control provided by the STM even allows mechanical gating experiments to be carried out, so that a
range of different transport regimes can be investigated for the same molecule \cite{ptcda1, ptcda2}. In general, both in its 
topographic and spectroscopical modes, an STM measurement consists of a collection of different $I$-$V$ curves for different 
tip-to-sample positions. 

In the vast majority of cases, calculations of STM currents are based on Bardeen's perturbative approach to
tunneling \cite{bar} and consist of evaluating the tunneling matrix elements between the STM tip and the sample. 
A simplified form due to Tersoff and Hamann (TH) \cite{th} which reduces to the calculation of the local density 
of states, is now mainstream and usually implemented in standard electronic structure codes. The TH scheme
assumes that the STM tip is sufficiently far from the molecule so as not to affect its electronic structure. As a result of
this assumption, there is no self-consistent evaluation of the potential drop nor of the charge re-distribution between 
the tip and the sample. These methods are thus not reliable when the tip is close to the molecule and deviation from the 
TH current are expected. It is therefore important to explore, for STM simulations, the use of self-consistent transport 
methods \cite{smeagol1,smeagol2,negf-dft} such as the non-equilibrium Green's function formalism (NEGF) \cite{negf} 
combined with {\it ab initio} electronic structures obtained from density functional theory (DFT) \cite{dft,ks}. This is the goal
of our work, which is based on the use of the electronic transport code {\sc smeagol} \cite{smeagol1,smeagol2}.

It is important to realize that NEGF-DFT is completely complementary to the standard TH scheme. Indeed, it
presents problems in the limit of large tip-to-sample separation, where TH is most effective. The main reason
for this is rooted in the fact that {\sc smeagol} employs the numerical implementation of DFT contained in the 
{\sc siesta} code \cite{siesta}. The {\sc siesta} basis set is constructed from numerical orbitals whose radial 
component is truncated beyond a certain cutoff radius \cite{siestabasis}, and extended vacuum regions are 
usually only poorly  described. In fact in the extreme case of a vacuum region extending well beyond all the 
cutoff radii, all matrix elements between the tip and the sample will vanish and the tunneling current will 
become identically zero. Improvements can be obtained by populating the vacuum region with empty 
orbitals (basis functions not associated to a pseudopotential). However for sensitive calculations of tiny tunneling 
currents the actual position of the empty orbitals can deeply affect the results, most typically by creating spurious 
current oscillations as a function of the tip-to-sample distance. Therefore we limit our method to tip-to-sample 
separation smaller than 6\AA\, for which reliable calculations do not need empty orbitals in the vacuum. 
We call this limit the quasi-contact limit.
\begin{figure}[ht!]
\begin{center}
\includegraphics[width=7.5cm,clip=true]{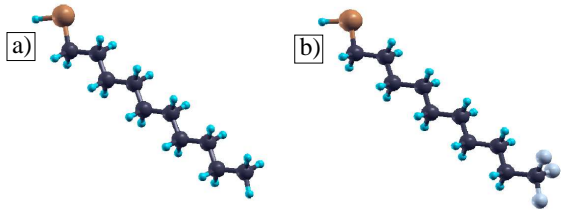}
\end{center}
\caption{\small{(Color on line) Decanethiol molecule with (a) $\mathrm{CH}_3$ (b) $\mathrm{CF}_3$ endgroup. 
Color code: C=black, S=brown, H=blue, F=purple.}}
\label{DectF}
\end{figure}

An example of experiments where the tip-molecule interactions are believed to be important, is that performed by 
Pflaum et. al. \cite{dtf}, in which a monolayer of alkanethiol molecules is deposited on a gold surface, and the transport 
properties are then probed by using an STM setup. The obtained zero-bias conductance is low, being of the order 
of $10^{-7} G_0$ at zero bias (the quantum conductance $G_0$ is defined as $2e^2/h$, with $e$ the electron charge
and $h$ the Planck constant). The $I$-$V$ curves are asymmetric and this asymmetry increases noticeably when the 
terminating CH$_3$ group is replaced by a CF$_3$ one. 
It is speculated that the additional asymmetry of the $\mathrm{CF}_3$-terminated molecule is due to a rearrangement 
of  the charge distribution near the end of the molecule caused by the high electronegativity of the F atoms. This in 
turn generates an electrostatic force between the STM tip and the molecule, the direction of which depends on the 
electrostatic potential at the tip. Such interaction causes the molecule to be either repelled or attracted by the tip
depending on the bias polarity, and creates the asymmetry in the $I$-$V$.

Calculations using the TH method have been performed for pentanethiol molecules on gold \cite{selloni}. 
However, the nature of the TH method requires a large tip-molecule separation, and the current obtained 
in these calculations is an order of magnitude smaller than that observed in the experiments. 
In this work, the mechanism behind the asymmetry in the $I$-$V$ curves of CH$_3$ and CF$_3$ terminated
alkanethiol molecules on gold is explored with NEGF-DFT. The issue related to the electrostatic interaction
between the tip and the sample is investigated in an approximate way. First we calculate from equilibrium 
DFT the electrical dipole on the molecules. Then we use classical electrostatic theory to determine the 
equilibrium position of the molecule with respect to the tip. Finally new transport calculations are carried out
for the newly estimated atomic positions. 

\section{Method: DFT-based Non-Equilibrium Green's Function Formalism}

\begin{figure}[ht]
\begin{center}
\includegraphics[width=6.5cm,clip=true]{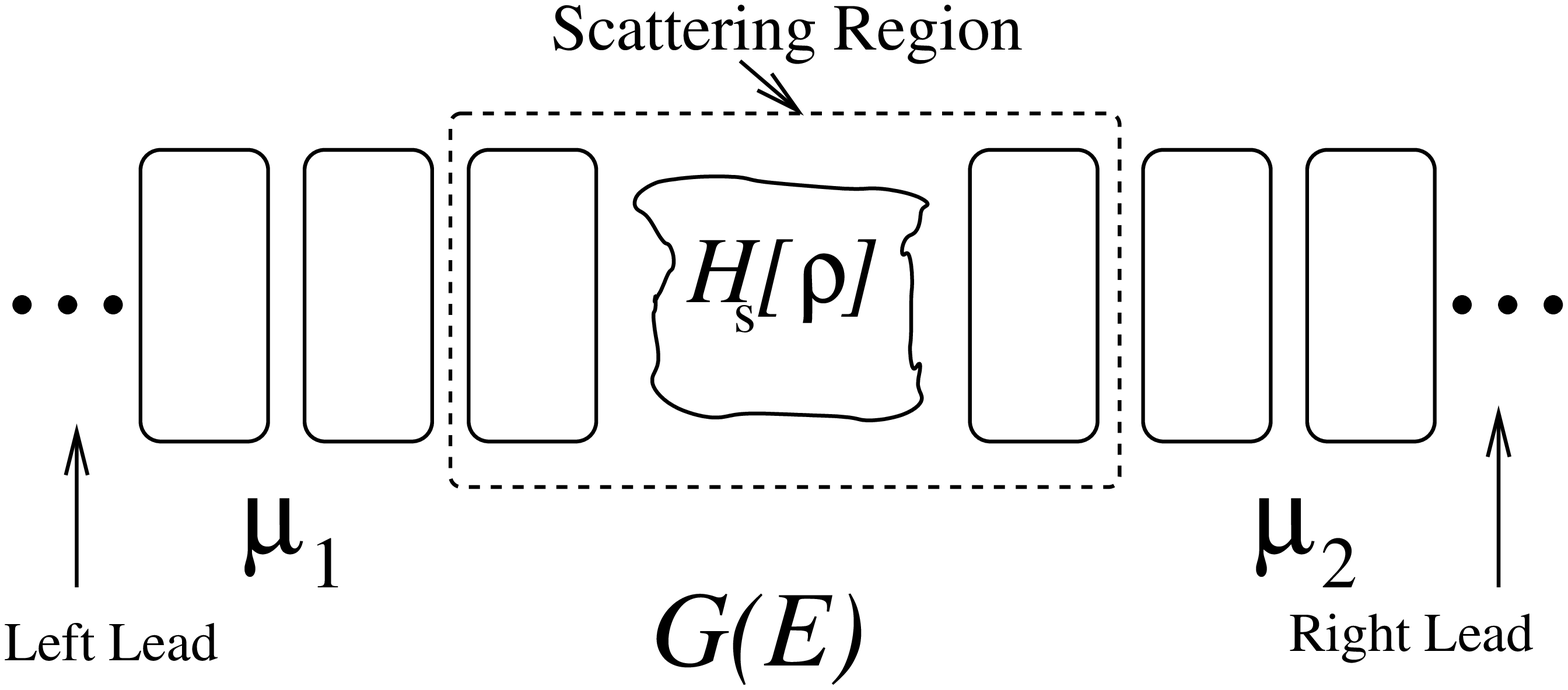}
\end{center}
\caption{\small{Schematic diagram of a metal-molecule-metal junction. A scattering region (the part of 
the system enclosed by the dashed box) is sandwiched between two current/voltage probes kept at the 
chemical potentials $\mu_\mathrm{1}$ and $\mu_\mathrm{2}$ respectively. The electrodes are modeled 
as being periodic in the direction of transport. A number of layers of the electrodes are included in the 
scattering region to allow the charge density to converge to the bulk value.}}
\label{Fig2}
\end{figure}

STM transport measurements are equivalent to calculating the two-probe $I$-$V$ curve of a molecule
sandwiched in between two metallic electrodes. One electrode represents the surface to which the molecule
is attached and the other is the STM tip. The NEGF scheme partitions such system into three regions, respectively 
the two current/voltage electrodes (leads) and a middle region called the scattering region (SR) (see figure \ref{Fig2}).

The SR includes both the molecule and a portion of the leads, and it is extended enough so that charge 
density calculated at the most external atomic layer resembles that of the bulk electrodes. The leads, which 
are assumed to be periodic crystals in the transport direction, are kept at different chemical potentials 
$\mu_\mathrm{1/2}=E_\mathrm{F}\pm eV/2$ where $V$ is the applied potential bias. We note this is somewhat 
different from STM experiments, where typically the chemical potential of the surface remains fixed while a 
potential bias is applied to the probe. However, such a difference in the definition does not effect the shape of 
the $I$-$V$ curve, since the final potential drop is calculated self-consistently. The SR is described by a 
Hamiltonian $H_\mathrm{s}$. This is used to construct the non-equilibrium Green's function

\begin{equation}
G(E)=\lim_{\eta\rightarrow 0}[(E+i\eta)-H_\mathrm{s}-\Sigma_\mathrm{1}-\Sigma_\mathrm{2}]^{-1}\:,
\label{negfmx}
\end{equation}
where $\Sigma_\mathrm{1/2}$ are the self-energies for the leads, constructed by semi-analytical derivation \cite{Ivan}
from the electronic structure of the bulk. $G(E)$ enters in a self-consistent procedure to calculate the density matrix, 
$\rho$, of the SR, and hence the two-probe current of the device \cite{negf,smeagol1,smeagol2,negf-dft,negftb}. 
The Hamiltonian for the scattering region $H_\mathrm{s}$ is generally assumed to be a function of the non-equilibrium 
charge density $\rho$, which is calculated following the NEGF prescription as
\begin{equation}
\rho = \frac{1}{2 \pi} \int^{\infty}_{-\infty} G(E)[\Gamma_\mathrm{1} f(E, \mu_\mathrm{1}) +  \Gamma_\mathrm{2} f(E, \mu_\mathrm{2})] G^{\dag}(E) dE \:,
\label{chdens}
\end{equation}
where $\Gamma_\mathrm{1/2} = i[\Sigma_\mathrm{1/2} - \Sigma^{\dag}_\mathrm{1/2}]$.
In practice, this integral is performed by splitting it into two parts\cite{negf,smeagol1,smeagol2,negf-dft,negftb}: 
an equilibrium part to be performed along a contour in the complex plane, and a non-equilibrium part 
to be evaluated along the real energy axis but that contributes only around $E_\mathrm{F}$. Finally, the converged 
Green's function can be used to calculate the two-probe current $I$ through the device
\begin{equation}
I = \frac{2e}{h} \int^{\infty}_{-\infty} \mathrm{Tr}[G(E )\Gamma_\mathrm{1} G^{\dag}(E) \Gamma_\mathrm{2}] (f(E, \mu_\mathrm{1}) -  f(E, \mu_\mathrm{2})) dE \:.
\label{curint}
\end{equation}
This is effectively the integral between $\mu_\mathrm{1}$ and $\mu_\mathrm{2}$ (the bias window) 
of the transmission coefficients $T(E) = \mathrm{Tr}[G(E )\Gamma_\mathrm{1} G^{\dag}(E) \Gamma_\mathrm{2}]$.

The NEGF scheme is general and not related to a particular electronic structure theory. In the case of {\sc smeagol}
\cite{smeagol1,smeagol2}, the electronic structure method used is density functional theory (DFT) \cite{dft} in the 
{\sc siesta} implementation. In particular in this work we will use the local density approximation \cite{ks} (LDA) 
of the exchange and correlation functional.

\section{Results}

\subsection{Electronic Structure of the Molecules}

The molecules used in the experiments of Pflaum et. al. are both methyl-terminated and fluorine-terminated 
alkanethiols. A range of different alkane chain lengths were used in that study, but most of the data were 
collected for decanethiols (see figure \ref{DectF}(a) and \ref{DectF}(b)). Decanethiol consists of an alkane chain 
of ten carbon atoms terminated with thiol (-SH) group. The sulphur atom in the thiol group forms a strong bond
with gold and anchors the molecule to the surface forming a well ordered self-assembled monolayer. 

\begin{figure}[ht!]
\begin{center}
\includegraphics[width=9.0cm,clip=true]{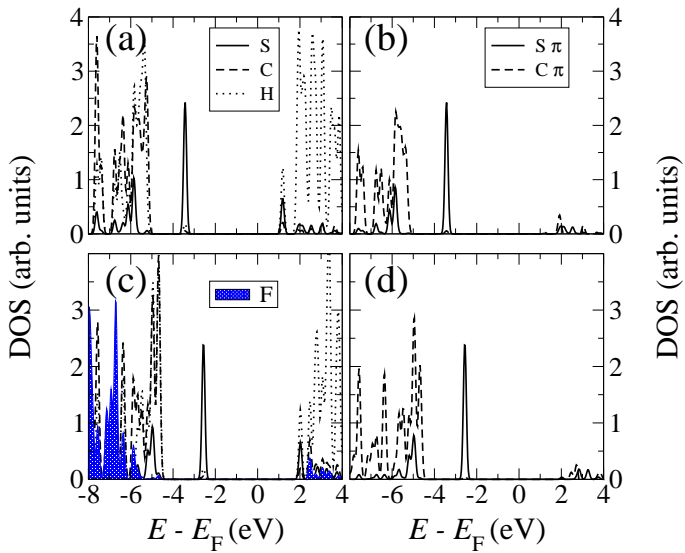}
\end{center}
\caption{\small{(Color on line) Orbital resolved DOS for the isolated decanethiol molecule. Panels (a) and (b)
are for the $\mathrm{CH}_3$ termination, while panels (c) and (d) are for the $\mathrm{CF}_3$.
In panel (b) and (d) only the DOS corresponding to $\pi$ orbitals perpendicular to the molecule
axis are displayed.}}
\label{Fig3}
\end{figure}

\begin{figure}[ht!]
\begin{center}
\includegraphics[width=8.0cm,clip=true]{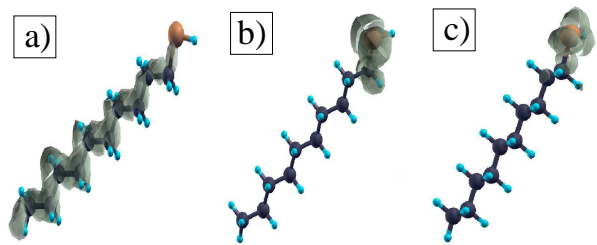}
\end{center}
\caption{\small{(Color on line) Local density of states for the isolated decanethiol molecule terminated 
with the $\mathrm{CH}_3$ group showing: (a) the HOMO-1 state (b) the HOMO and (c) the LUMO. The local 
DOS for the molecule terminated with the $\mathrm{CF}_3$ group is similar and it is not displayed here. 
Color code: C=black, S=brown, H=blue, F=purple.}}
\label{Fig4}
\end{figure}

The orbital resolved density of states (DOS) for isolated decanethiol both $\mathrm{CH}_3$-
and CF$_3$-terminated is shown in figure \ref{Fig3}. The energy gap between the highest
occupied molecular orbital (HOMO) and the lowest unoccupied molecular orbital (LUMO) is
in both cases rather large ($\sim$~5~eV ) \cite{note1}. Most importantly the DOS of the two frontier molecular 
orbitals do not have large amplitude over the C atoms forming the decanethiol, but are rather
localized on the $\pi$ states of S perpendicular to the molecule axis (see figures \ref{Fig3}(b)
and \ref{Fig3}(d)). In the case of CF$_3$-termination the F contribution to the DOS is only
confined to part of the LUMO and to low energy HOMO levels. The orbital nature of the HOMO and
LUMO is further investigated in figure \ref{Fig4} where we present the local DOS for the CH$_3$-terminated 
molecule (the one for the CF$_3$-terminated case is rather similar and it is not presented here).

We note that both HOMO and LUMO have amplitude mainly around the thiol group with little
charge spread over the alkane chain. One should therefore expect little conductance through 
those states. In contrast the first state below the HOMO (HOMO-1) (figure \ref{Fig4}(a)) appears as a 
delocalized $\pi$ state and it is expected to conduct efficiently.

\subsection{Transport Properties of the Molecules on Gold Surface}

\begin{figure}[ht]
\begin{center}
\includegraphics[width=9.0cm,clip=true]{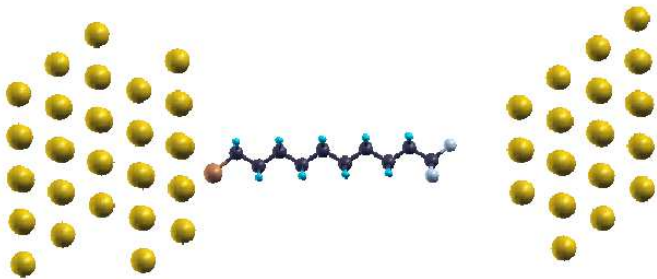}
\end{center}
\caption{\small{(Color on line) Simulation cell used in the transport calculations: a $\mathrm{CF}_3$-terminated 
decanethiol molecule is attached to gold surface and probed with a Au STM tip. The case of the 
$\mathrm{CH}_3$-terminated molecule is similar and it is not displayed here. Color code: Au=yellow, C=black, 
S=brown, H=blue, F=purple.}}
\label{Au_DTF}
\end{figure}

In order to calculate the transport properties, we attach the molecules to the Au fcc (111) hollow site
via a terminating thiol group at a sulphur-surface distance of 1.9~\AA \cite{AuSdistDef}. This is the 
equilibrium distance measured for the hollow site configuration \cite{AuSdist}. The arrangement of the molecule 
on the surface and the relative position of the tip is shown in figure \ref{Au_DTF} for the $\mathrm{CF}_3$-terminated 
molecule. The configuration for the $\mathrm{CH}_3$-terminated case is similar and it is not displayed here.
In the experiments from Pflaum et. al. $\mathrm{CH}_3$-terminated decanethiols are tilted at an angle of $\sim 32^{\circ}$
from the vertical direction, while for $\mathrm{CF}_3$-terminated the tilting angle is $\sim 39^{\circ}$ \cite{dtf}.
Our simulation cells however have the molecules placed perpendicular to the surface. This allows the unit cell used 
to be smaller than that required for simulating such tilting angles, and the computational overheads are greatly reduced. 
We also make some approximations in modeling
the STM tip. First of all we consider a gold tip instead of the used tungsten or platinum-iridium ones, in such a way 
to avoid issues related to having two electrodes with different work functions. In doing this we describe the electronic 
structure of the electrodes with a $6s$ only basis set, which was proved to provide a reasonably accurate choice for
transport calculations at only a moderate computational cost \cite{bdtsic}. Secondly we do not consider an
atomically sharp tip but a slightly blunter one. This geometry improves the stability of the calculations, allowing 
the self-consistent convergence to be quicker and more reliable. 

\begin{figure}[ht!]
\begin{center}
\includegraphics[width=9.5cm,clip=true]{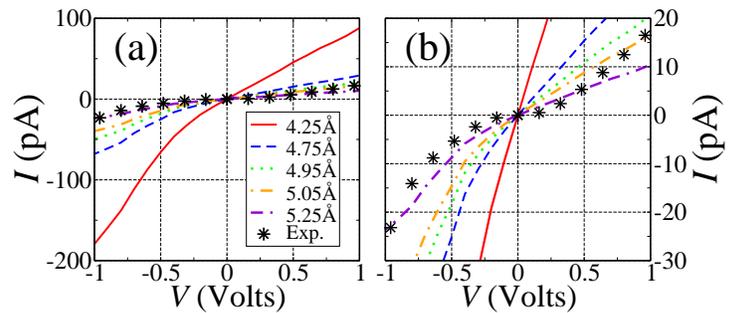}
\end{center}
\caption{\small{(Color on line) $I$-$V$ curves for CH$_3$-terminated decanethiol attached to a gold surface for 
different distances between the C atom in the CH$_3$ group and the probing tip. Changing this distance by 
0.1~\AA~causes the size of the current to change by approximately one order of magnitude at 1~V. Panel (b) is 
a zoom of panel (a) in the current range between [-30,20]~pA.}}
\label{curdist}
\end{figure}

As expected the tunneling current is extremely sensitive to the distance between the tip and the molecule, so that
it is rather difficult to determine from experiments the precise tip vertical position. Our working strategy is then to 
adjust this distance to match the magnitude of the current obtained in experiments. Such a a tuning exercise is
presented in figure~\ref{curdist} where we show the $I$-$V$ curves for different tip-to-molecule separations, where
these are defined as the distance between the terminating surface of the tip and the C atom in the CH$_3$
group. From the figure one can deduce that the best match is obtained for a distance of 5.25\AA. The results presented 
in the rest of this section are obtained for this tip-to-molecule separation.

\begin{figure}[ht]
\begin{center}
\includegraphics[width=9.0cm,clip=true]{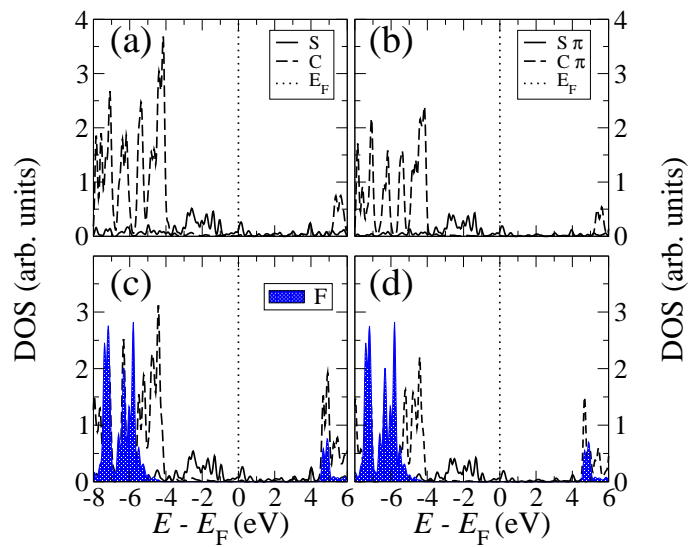}
\end{center}
\caption{\small{(Color on line) Orbital resolved DOS for the decanethiol molecule attached to the 
gold surface. Panels (a) and (b) are for the $\mathrm{CH}_3$ termination, while panels (c) and (d) are 
for the $\mathrm{CF}_3$. In panel (b) and (d) only the DOS corresponding to $\pi$ orbitals perpendicular 
to the molecule axis are displayed for C and S, while in panel (d) all the fluorine $\pi$ orbitals are 
shown. The HOMO-LUMO gap is large, although the HOMO is about 1~eV below $E_\mathrm{F}$.}}
\label{AuDTpdos}
\end{figure}

\begin{figure}[ht]
\begin{center}
\includegraphics[width=7.5cm,clip=true]{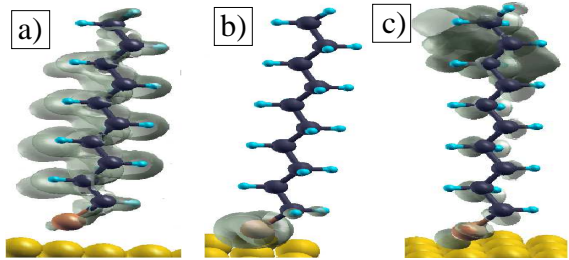}
\end{center}
\caption{\small{(Color on line) Local DOS isosurface for the $\mathrm{CH}_3$-terminated decanethiol molecule on 
the Au (111) surface: (a) the HOMO-1 level, (b) the HOMO level and (c) the LUMO level. Note how the HOMO 
is localized around the thiol endgroup and it is not expected to conduct efficiently.  Color code: Au=yellow, C=black, 
S=brown, H=blue.}}
\label{AuDTldos}
\end{figure}

\begin{figure}[ht]
\begin{center}
\includegraphics[width=7.5cm,clip=true]{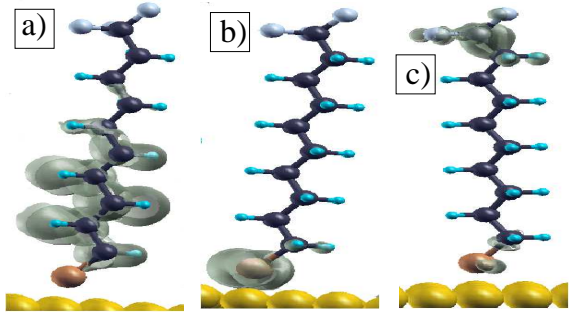}
\end{center}
\caption{\small{(Color on line) Local DOS isosurface for the $\mathrm{CF}_3$-terminated decanethiol molecule on 
the Au (111) surface: (a) the HOMO-1 level, (b) the HOMO level and (c) the LUMO level. Note how the HOMO 
is localized around the thiol endgroup and it is not expected to conduct efficiently.  Color code: Au=yellow, C=black, 
S=brown, H=blue., F=light blue.}}
\label{AuDTFldos}
\end{figure}

The orbital resolved DOS for both $\mathrm{CH}_3$- and $\mathrm{CF}_3$-terminated decanethiol on Au are 
shown in figure \ref{AuDTpdos}. In general and for both the molecule the HOMO is relatively close to the 
gold $E_\mathrm{F}$. 
However, a closer look at its local DOS (this is calculated as the local DOS over an energy window 
corresponding to the HOMO level) depicted in figures \ref{AuDTldos} and \ref{AuDTFldos}, reveals that
such a state is mainly localized around the S atoms of the thiol groups. The delocalized 
$\pi$-orbitals capable of efficient charge transport across the molecule are further away from the Fermi level
(approximately 4~eV for both molecules). It is also interesting to observe the effects that the different 
terminating groups (CH$_3$ and CF$_3$) have over the electronic structure of the HOMO-1 level. The most
notable difference is associated to the charge density distribution along the CH$_2$ groups 
forming the decanethiol. While in the case of CH$_3$ termination the electron charge distributes almost 
uniformly over CH$_2$ up to near the terminating CH$_3$ group, in the case of CF$_3$-terminated molecule there 
is a strong modulation with a decay of the charge density as the CF$_3$ group is approached. This
is due to the Coulomb repulsion from the more electronegative CF$_3$ group. Such an
electrostatic driven charge decay is reminiscent of the same effect investigated by STM for molecules on surfaces 
in the presence of point charges \cite{Wolcok}.

\begin{figure}[ht!]
\begin{center}
\includegraphics[width=9.0cm,clip=true]{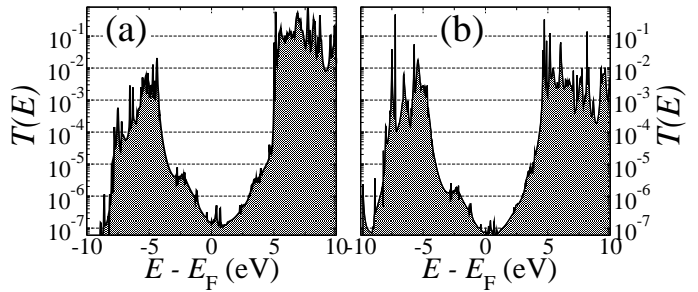}
\end{center}
\caption{\small{Transmission coefficients at zero bias for decanethiol attached to gold surface and terminated
with (a) $\mathrm{CH}_3$ and (b) $\mathrm{CF}_3$ endgroup. Note the gap in the transmission of 
about 5~eV on either side of the Fermi level.}}
\label{transmdt_dtf}
\end{figure}

\begin{figure}[ht!]
\begin{center}
\includegraphics[width=9.0cm,clip=true]{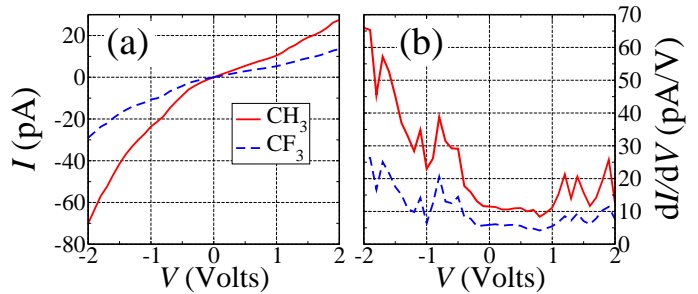}
\end{center}
\caption{\small{(Color on line)
(a) $I$-$V$ curve and (b) differential conductance for decanethiol molecules terminated with 
both $\mathrm{CH}_3$ and $\mathrm{CF}_3$ groups. Note the bias asymmetry, with the conductance at negative 
bias being 2-3 times larger than that for positive bias. The conductance is lower for the $\mathrm{CF}_3$-terminated
molecule than that of the $\mathrm{CH}_3$-terminated.}}
\label{curdfcl20}
\end{figure}

Despite these differences the absence of any extended molecular orbital near the Fermi level of
gold is common to both molecules and therefore we expect just a tiny tunnel current at low bias. 
This can be appreciated by looking at both the transmission coefficients, $T(E)$ (figure \ref{transmdt_dtf})
and the $I$-$V$ curves (figure \ref{curdfcl20}). There is a large gap in the resonances in the zero 
bias transmission coefficients of about 5~eV on each side of the Fermi level with the edges of such a 
gap corresponding to the HOMO-1 and the LUMO, respectively below and above the Au Fermi energy.
The transmission of the HOMO provides only a small shoulder in $T(E)$ ($T\sim 10^{-6}$) at about
2.5~eV below $E_\mathrm{F}$, confirming that the HOMO does not provide an efficient transport channel 
at resonance. However, since the HOMO is the first molecular level to enter the bias window (at least
for one of the two current polarities), it is expected to dominate the low-bias region of the $I$-$V$.

Moving our attention to the $I$-$V$ curves and the differential conductances shown in figure \ref{curdfcl20},
the most noticeable feature is their asymmetry. In particular conductance at positive bias is about between 
2 to 3 times smaller than that for negative bias. Moreover we observe that the conductance of CH$_3$-terminated 
molecules is both larger and more asymmetric of that for CF$_3$-terminated. 

\begin{figure}[ht!]
\begin{center}
\includegraphics[width=9.0cm,clip=true]{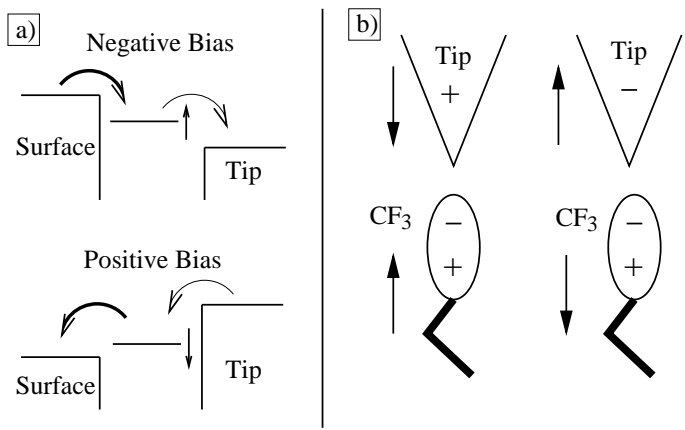}
\end{center}
\caption{\small{(a) Schematic energy level diagram of the STM geometry. Since the molecule is more strongly coupled 
to the surface than to the tip its occupation is determined by the surface chemical potential, i.e. it will charge for negative
bias and de-charge for positive. Such a charging mechanism produces a shift of the occupied molecular level towards
higher (lower) energy for negative (positive) bias.
(b) Cartoon showing the electrostatic interaction between tip and molecule $\mathrm{CF}_3$ endgroup. 
In this case there is an electric dipole forming, so that the molecule will be attracted or repulsed by the STM tip, 
depending on the bias polarity.}}
\label{tipforce}
\end{figure}

The reason for the conductance asymmetry is rooted in the different electronic coupling strength between the 
molecule and the gold surface as compared to the much weaker one between the molecule and the STM tip. 
The S atom in fact forms quite a strong bond with the gold surface, whereas the tip is separated by a vacuum
region which provides a substantial potential barrier. Such difference in coupling persists to rather close tip-to-molecule
separations since both the CH$_3$ and CF$_3$ endgroups do not bind to the Au tip.
The mechanism leading to the asymmetry is then illustrated in the cartoon of figure~\ref{tipforce}(a). 
Since the different coupling strengths of the two electrodes with the molecule, electrons are easily transferred from 
the substrate to the molecule, but they can hardly hop to the tip. This means that the occupation of the molecule 
is determined by the chemical potential of the substrate and that the molecule gains electronic charges 
when the electron flux is from the surface to the tip (negative bias according to our definition). In contrast the molecules 
lose charge when the flux is in the other direction (positive bias). Since charging produces a shift of the molecular 
electronic states towards higher energies one expects the HOMO to move into the bias window at negative voltages 
and away from it at positive ones. The same behaviour is expected for the HOMO-1, while the LUMO
follows exactly the opposite pattern (it moves into the bias window for positive bias and away from it for negative)
\cite{NoteBias}. 

\begin{figure}[ht!]
\begin{center}
\includegraphics[width=9.0cm,clip=true]{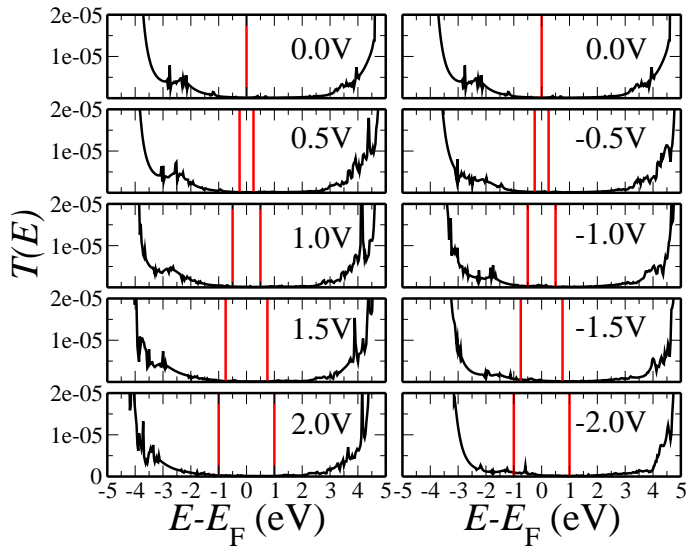}
\end{center}
\caption{\small{Transmission coefficients as a function of energy for different biases for $\mathrm{CH}_3$-terminated
decanethiol. Note how the resonances in the transmission coefficients due to the occupied states move up in energy 
closer to the bias window at negative bias, and move away from the bias window for positive bias. The vertical lines 
mark the bias window.}}
\label{transmdtl20}
\end{figure}

\begin{figure}[ht!]
\begin{center}
\includegraphics[width=9.0cm,clip=true]{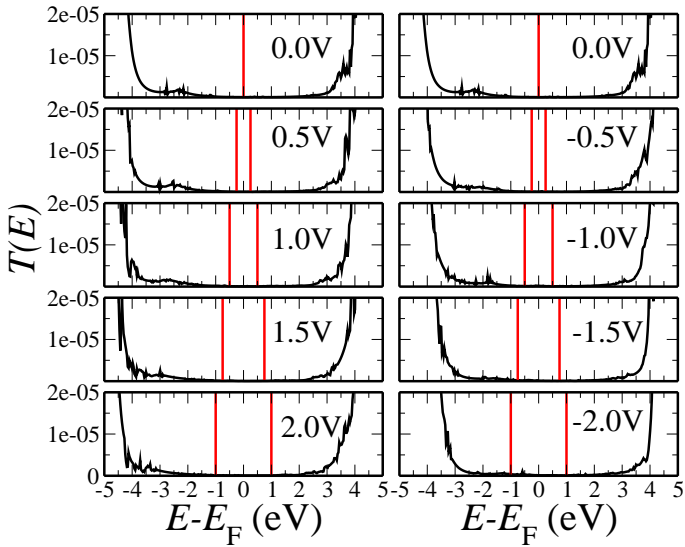}
\end{center}
\caption{\small{Transmission coefficients as a function of energy for different biases for $\mathrm{CF}_3$-terminated
decanethiol. Note how the resonances in the transmission coefficients due to the occupied states move up in energy 
closer to the bias window at negative bias and move away from the bias window for positive bias. The vertical lines 
mark the bias window.}}
\label{transmdtfl20}
\end{figure}

The electrostatic evolution of the system under bias is illustrated in figures \ref{transmdtl20} and \ref{transmdtfl20}
where we show the transmission coefficients as a function of energy for different voltages. Clearly the behaviour
is the one expected by our simple electrostatic model with an upshift (downshift) in energy of the molecular
levels at negative (positive) bias. Since the highly conducting HOMO-1 and LUMO levels never enters the bias
window in this voltage range (limited to $\pm$2~Volt), the onset of the current is solely determined by the position 
of the HOMO. This however can enter the bias window only for negative bias explaining the current asymmetry. 
Furthermore, since the transmission of the HOMO for CH$_3$-terminated decanethiol is larger than that of the
CF$_3$-terminated case, the conductance of CH$_3$-decanethiol is both larger and more asymmetric than 
that of its CF$_3$ terminated counterpart. 

\begin{figure}[ht!]
\begin{center}
\includegraphics[width=9.0cm,clip=true]{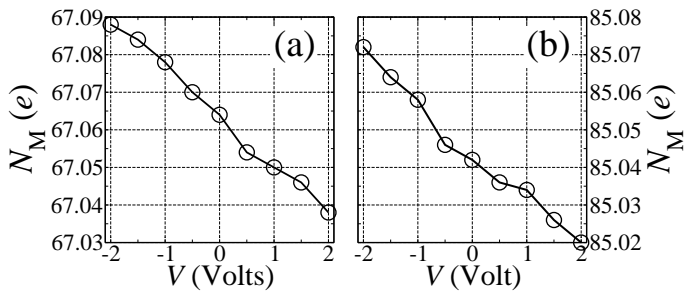}
\end{center}
\caption{\small{Mulliken populations for decanethiol molecule terminate with (a) $\mathrm{CH}_3$ and 
(b) $\mathrm{CF}_3$ groups and attached to the gold surface. Note how in both cases the occupation of the 
molecule drops as the bias increases from negative to positive.}}
\label{mulldtfl}
\end{figure}

Finally, in figure~\ref{mulldtfl} we present the total M\"ulliken population, $N_\mathrm{M}$, of the molecule as a function of 
voltage. The figure provides quantitative evidence for the charging of the molecule as a function of bias and one can observe
a total charge variation of about 0.05~$e$ over the entire bias range of 4~Volt. Interestingly, and despite the different terminating
groups, the charge variation with bias is essentially identical for the two terminations.

\subsection{Influence of the tip-to-molecule electrostatic interaction over the transport}

As concluded in the previous section, the asymmetry of the conductance with respect to the bias direction found in 
our calculations is due to the asymmetry in the electronic coupling of the molecule to the leads. Interestingly a strong 
asymmetry was also observed in the experimental conductance measurements for the same molecules \cite{dtf}.
However, in the experiments, $\mathrm{CF}_3$-terminated molecules displayed a far more pronounced asymmetry
than the $\mathrm{CH}_3$-terminated ones. This is in contrast with our findings where the asymmetry is similar, and 
in fact it is more pronounced for the $\mathrm{CH}_3$-terminated case. In agreement with the initial suggestion from
Pflaum et al., here we argue that this additional asymmetry is not of electronic origin but rather is connected to 
the electrostatic interaction between the tip and the molecule under bias. This re-positions the molecule with respect to
the tip, effectively changing the tunneling current. 

\begin{figure}[ht!]
\begin{center}
\includegraphics[width=9.0cm,clip=true]{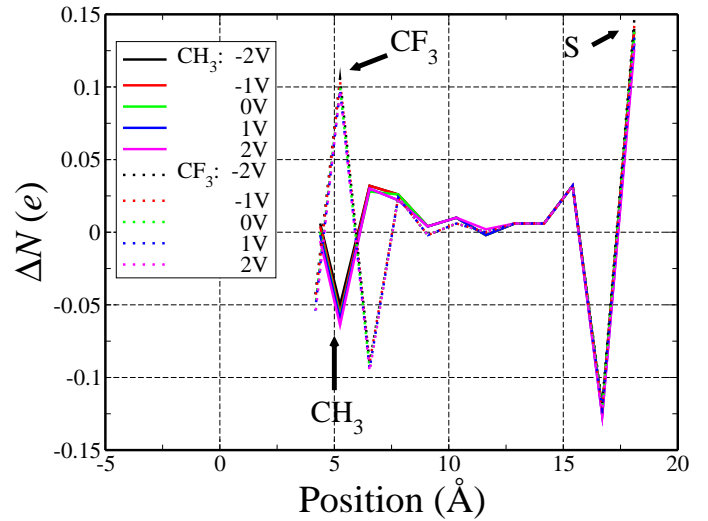}
\end{center}
\caption{\small{(Color on line) Net occupation of the atoms as the function of position along the axis of the molecule.}}
\label{ncdtfl20}
\end{figure}

The main idea is illustrated schematically in figure \ref{tipforce}(b). The large electronegativity of the F atoms in the CF$_3$
group is expected to displace a substantial net charge towards the end of the molecule. This then interacts electrostatically
with the surface charge of the STM tip, either repelling or attracting the molecule depending on the bias polarity. Such a charge
displacement is illustrated in in figure~\ref{ncdtfl20} where we show the charge excess with respect to neutrality, $\Delta N$,
as calculated from the M\"ulliken population as a function of the position along the molecule axis. In our notation $\Delta N>0$
($\Delta N<0$) indicates a negatively (positively) charged portion of the molecule. Substituting H by F atoms 
causes an increase in the occupation of the endmost carbon atom, although the F atoms show a net positive charge. 
Overall the total net charge on the $\mathrm{CF}_3$ group is negative (it posses extra electrons), whereas the total net 
charge on the $\mathrm{CH}_3$ group is positive. In addition the two end groups produce an electrical dipole
with opposite sign and different magnitude, with that associated to CF$_3$ being the largest. Finally it is important to
observe that by large the charge distribution over the molecule is not affected by the external electric field so that $\Delta N$
is almost bias independent for both the terminations.

We estimate the forces acting on the molecule by using simple classical electrostatic theory. In the case of
CH$_3$-terminated decanethiol we calculate a force of $3 \times 10^{-3}$ eV \AA$^{-1}$ in the direction of 
the tip at the positive bias of 2~Volt, and $2 \times 10^{-3}$ eV \AA$^{-1}$ in the direction of the substrate 
at the negative bias of 2~Volt. These have to be put in relation with the elastic forces needed to stretch the
molecule. DFT calculations performed using {\sc siesta} demonstrate that the tip-to-molecule electrostatic force 
at positive bias causes the molecule to stretch by about 0.002~\AA, while at negative bias the compression is
of the order of 0.001~\AA~(at $\pm2$~Volt).
Similarly, the force acting on the $\mathrm{CF}_3$ group is calculated to be of the order of 
$3 \times 10^{-3}$ eV \AA$^{-1}$ in the direction of the substrate when a positive bias of 2~Volt is applied, 
and of the order of $4 \times 10^{-3}$ eV \AA$^{-1}$ in the direction of the tip at a negative bias of 2~Volt. 
These forces are in the opposite direction to those on the $\mathrm{CH}_3$ termination due to the 
opposite directions of the two dipoles. Again DFT calculations for $\mathrm{CF}_3$-decanethiol indicate 
that the force at positive bias causes the molecule to compress by about 0.001~\AA, and the force at negative 
bias to stretch by 0.002~\AA. These length changes are of the same magnitude for both types of endgroup, 
but they occur at opposite bias directions due to the opposite direction of the electrostatic force in the two cases.

The stretching (or compressing) of the molecule under bias alters the tip-to-molecule separation, which in turn 
affects the magnitude of the current. Let us look at the CH$_3$ case first. At positive bias the molecule is stretched
so that its end is closer to the STM tip, the tip-to-molecule separation is reduced and the current will increase. 
In contrast at negative bias, the molecule is compressed and a reduction of the current is expected. Since the 
current obtained from our calculations at negative bias is larger than that for positive bias, such a change in molecule 
length would contribute to reducing the $I$-$V$ asymmetry in agreement with the experimental results.
However, a quantitative estimate of the changes in the $I$-$V$ curve as a function of the molecule compression/expansion
reveals that molecular displacements of this tiny magnitude (a few $\%$ of an \AA) are insufficient to make a noticeable 
change to the $I$-$V$.  

The results for $\mathrm{CF}_3$-terminated decanethiol are similar. At positive bias, the molecule is compressed
so that the $\mathrm{CF}_3$ group is further away from the STM tip, while at negative bias the molecule is stretched 
and the tip-to-molecule separation is reduced. Since the current calculated at negative bias is larger than that at 
positive bias, the electrostatic interaction in this case has the effect of enhancing the asymmetry of the $I$-$V$ curves. 
This is also consistent with the experimental results, showing a larger asymmetry in the transport for the molecule 
terminated with the $\mathrm{CF}_3$ group. However, similar to the $\mathrm{CH}_3$-terminated case, the estimated
molecule length change is insufficient to make a noticeable modification to the $I$-$V$ curve. Note that these
negative estimates are all obtained by comparing the geometry at zero bias with that corresponding to the largest 
displacement (at $\pm 2$~Volt), so that they should be considered as the upper bound for the current changes. 
Therefore we can safely conclude that changes in molecule length can hardly be at the origin of the experimentally
observed asymmetry in the $I$-$V$ curve. 

\begin{figure}[ht!]
\begin{center}
\includegraphics[width=5.0cm,clip=true]{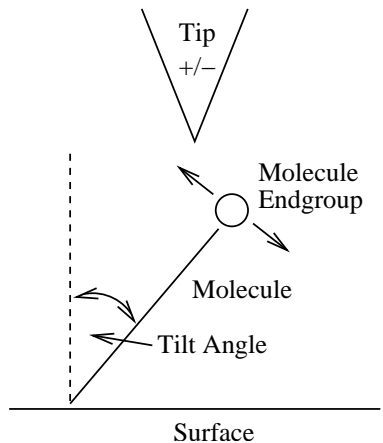}
\end{center}
\caption{\small{Schematic illustration showing the tilting angle of the molecule over the Au surface and the rotation due to 
the interaction with the STM tip.}}
\label{molangforce}
\end{figure}

A further possibility is to consider changes in the bonding angle between the molecule and the substrate. As mentioned 
before, decanethiols do not arrange vertically to the surface, but they rather form angles of $32^{\circ}$ and
$39^{\circ}$, respectively for the $\mathrm{CH}_3$ and $\mathrm{CF}_3$ terminations \cite{dtf}. Hence it is likely 
that the interaction between the STM tip and the endgroups causes the molecule to rotate either away from or towards the 
STM tip, as shown in figure \ref{molangforce}. The direction of the rotation depends again on the sign of the electrostatic
force. However since the energy required to change the bonding angle is expected to be rather smaller than that
required for an axial distortion, we expect changes in bond angle to produce larger changes in the tip-to-molecule
separation. 

\begin{figure}[ht!]
\begin{center}
\includegraphics[width=9.0cm,clip=true]{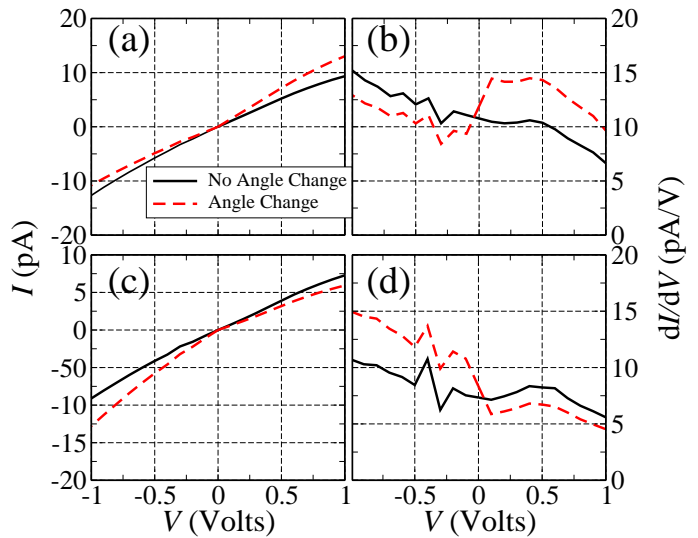}
\end{center}
\caption{\small{(Color on line) $I$-$V$ curves and differential conductance as a function of bias obtained both with 
and without considering the effects of the electrostatic-induced rotation of the molecule. Panels (a) and (b) show the 
$I$-$V$ and $\mathrm{d}I/\mathrm{d}V$ curves for the $\mathrm{CH}_3$ termination, and (c) and (d) show the same 
for the $\mathrm{CF}_3$ case.}}
\label{curangdtdtf}
\end{figure}

Using the same electrostatic forces calculated before we estimate for the $\mathrm{CH}_3$ termination 
a $1^{\circ}$ rotation towards the STM tip at positive bias (2~Volt), and a $0.5^{\circ}$ rotation towards the
surface at negative bias (-2~Volt). Similarly for the $\mathrm{CF}_3$ termination, the rotations are along the opposite 
direction with changes in the bond angle of  $0.5^{\circ}$ towards the surface at positive bias and $1^{\circ}$ 
towards the STM tip at negative bias. These rotations are sufficient to alter the $I$-$V$ characteristics, as shown for 
both types of molecule in figure \ref{curangdtdtf}.

In the case of $\mathrm{CH}_3$ termination the calculated rotation is actually sufficient to suppress the 
asymmetry in the $I$-$V$ present because of the asymmetric coupling to the electrodes. The current at 1~Volt is now 
larger for positive bias than that at the same negative bias. In contrast for $\mathrm{CF}_3$-terminated 
molecules, the rotation is such that the $I$-$V$ asymmetry already present is enhanced. Thus, when the 
electrostatic-induced rotation of the molecule is taken into account, the $I$-$V$ curve for the 
$\mathrm{CF}_3$-terminated molecule appears as more pronounced than that obtained for the 
$\mathrm{CH}_3$-terminated one, in semi-quantitative agreement with experiments \cite{dtf}.  

In concluding we wish to point out a few possible sources of error in our calculations, which might affect
a fully quantitative comparison with experiments. Firstly we remark that the Au atoms of both the tip and
the substrate are described at the 6$s$-only level. This provides a good description of the
Au Fermi surface, thus of the transport properties, but usually it does not provide accurate interatomic forces. 
Secondly the calculated forces between the molecule and the tip certainly depend on the actual shape and 
detailed position of the tip. These are difficult to characterize with certainty and errors are therefore difficult to avoid. 
Finally the actual value of the tilting angle depends on the inter-molecular interaction is the self-assembled 
mono-layer, and thus the force required to change the angle would depend on the details of the 
molecular coverage (density, symmetry, bonding site, etc.). 

\section{Conclusion}

We have demonstrated that the DFT-NEGF code {\sc smeagol} can be used to simulate STM-type 
experiments in the near to contact regime. This is under the working condition of the tip-to-sample 
distance being sufficiently small that the basis orbitals have not been artificially cut off (i.e. the vacuum 
region between the tip and the surface is still well described). {\sc smeagol} then allows us to investigate 
systems in which the tip interacts with the molecule, and to study effects of finite bias on the electronic structure 
of the molecule.

Calculations for alkanethiol molecules with STM-type arrangements show strong asymmetry in the 
$I$-$V$ curves, which can be explained by the asymmetry in the coupling to the two different electrodes
(substrate and tip). However, the calculated asymmetry is similar for both $\mathrm{CH}_3$- and 
$\mathrm{CF}_3$-terminated decanethiols, in contrast to the experimental measurements, showing a far
stronger asymmetry for the case of $\mathrm{CF}_3$ termination. 
We have then investigated the original suggestion, which attributed the asymmetry to small configurational 
changes of the molecule under bias due to electrostatic interaction between the tip and the substrate. 
Our calculations demonstrate that changes to molecule length are too small and unlikely to have any major 
impact on the $I$-$V$ curves, however a rotation of the molecule may cause significant changes to the current. 

As a final remark we point out that the results presented in this work have obtained by using the LDA. 
However, LDA has several shortfalls 
which can strongly affect electron transport calculations. In this case, approximate self-interaction corrections 
such as ASIC \cite{dasc, bdtsic} are unlikely to offer a substantial improvement, since the conductance is 
due to off-resonance tunneling, and would not be particularly sensitive to the exact  position of the molecular 
orbitals. However, for these molecules, an accurate calculation of the electric polarisability beyond the LDA might
be important \cite{ASICPOL}.

\section*{Acknowledgments}

We thank G.~Scoles for having driven our attention to the experiments discussed in this paper. 
This work is funded by Science Foundation of Ireland (grant 07/IN.1/I945). Computational resources have 
been provided by the HEA IITAC project managed by the Trinity Center for High Performance Computing and by
ICHEC.
 \\


\begin{thebibliography}{99}
\bibitem{note1}Note that this is the LDA HOMO-LUMO gap calculated from the Kohn-Sham eigenvalues. As such, in general 
it is not expected to be a good  estimate of the actual experimental gap, which is the difference between
the ionization potential and the electron affinity of the molecule. In this case however the junction is in a tunneling 
regime already at this LDA level and possible corrections are expected to produce only minor quantitative changes to
the calculated transport properties.
\bibitem{AuSdistDef}The sulphur-surface distance is defined as the distance between the S atom of the thiol group
and the hollow position in the fcc (111) surface.
\bibitem{NoteBias}Note that in our transport calculation the bias window is introduced by shifting the chemical potentials
of both the tip and the substrate by $\pm eV/2$. This is however completely equivalent to the standard STM setup where the
sample potential is kept fixed and the voltage is applied to the tip. Note also that the definition of bias polarity in our work is
opposite to that of Pflaum et al. \cite{dtf}.
\end{thebibliography}
\end{document}